\documentclass{aipproc}
\layoutstyle{6x9}
\usepackage{amssymb,latexsym}
\usepackage{graphics}
\usepackage{amsmath}
\begin{document}
\title{Schr\"odinger Equation for Joint Bidirectional Evolution in Time:
Astrophysical Applications}
\author{G. E. Hahne}{
        address={P. O. Box 2748, Sunnyvale, CA 94087},
          altaddress={email:  ghahne@mail.arc.nasa.gov}}
\begin{abstract}
      A straightforward extension of quantum mechanics 
and quantum field theory is proposed
that can describe physical systems comprising two
interacting subsystems:  one subsystem
evolves forward in time, the other, backward.
The space of quantum states is the direct sum of the states
representing the respective subsystems, whereupon
there are two linearly independent vacuum states, 
one each for the forward and the backward subspace.
An indefinite metric is
imposed on the space of quantum states 
such that purely forward (respectively, purely backward)
states have positive (respectively, negative) norms.
Hamiltonians are self-adjoint operators with respect
to the metric, such that
interactions/transitions between the subspaces can be accounted for.
Given suitable definitions of input and of output
states at the two ends of a time interval, input and output states
separately have positive norms such that
probability is conserved, and hence $S$-matrices are unitary.
A discussion of the physics entailed in the proposed
formalism is undertaken.  Then as an application, a simple model 
of a relativistic quantum field theory is proposed.
In this model, the expected vacuum energy (thought to be
associated with the
cosmological constant) almost vanishes uniformly for times
in an interval due to cancellation of the energies
of the forward and backward vacuum states;  this cancellation holds
whatever be the input vacuum state at the ends of the time interval. 
A model is advanced wherein magnetic monopoles live exclusively
in backward-evolving states, and interact with forward-evolving
electric charges in a certain way.
Proposals for further research, particularly concerning the 
possible detection
of advanced gravitational waves, and a conjecture on the
physics of dark matter and dark energy, conclude the report.
\end{abstract}
\maketitle
\section{1. Introduction}

     This report comprises a sketch of, and an elaboration of,
material published previously \cite{R:Hahne1}.
In the the present Sec.~1, we shall review the construction
proposed in \cite{R:Hahne1} for an extension of quantum mechanics that
can describe physical systems in which both forward
and reverse causality in time can occur jointly
and interactively.

     Let $\mathcal{H}^F$ be the Hilbert space incorporating
quantum states of the contents of the
known world, comprising the quantum fields
describing all possible states and occupations of 
electrons and other leptons, protons and other hadrons, 
photons and other vector bosons, etc.  (Note: we shall exclude gravitation
from the quantum world, and treat the gravitational field as governed
by classical field equations, which determine the structure
of the background space-time within which quantum phenomena evolve.)
This world is known
to evolve forward (hence the superscript $F$) in time; that is, 
given the physical state at an earlier time one can predict
the probabilities of various outcomes at a later time.
Now suppose that there is another Hilbert space $\mathcal{H}^B$,
where the ``$B$'' stands for ``backward'', wherein
physical phenomena evolve contrariwise, that is, given the
quantum state of a system at a later time, one can predict the
probability of various outcomes at an earlier time.
We can call the latter a world ``antiparallel'' to our own.
An intuitive, geometrical picture of the system comprises two
parallel Minkowski spacetimes identified point-by-point
in a metric-preserving way, such that certain interactions,
or quantum jumps, can occur between particles living on different
subspaces, but such that the dynamical time evolution in
the two spacetimes is oppositely directed.  The notion of these
parallel spaces being embedded in some higher-dimensional
spacetime is superfluous, and indeed is inconsistent with their 
oppositely directed causality in time.

     Suppose further that the universe of quantum states
comprises the direct sum of the above two spaces,
\begin{equation}\label{E:1}
\mathcal{H}\ =\ \mathcal{H}^F\,\oplus\,\mathcal{H}^B.
\end{equation}
This means that if $\Psi^F$ is a state in $\mathcal{H}^F$
and $\Psi^B$ is a state in $\mathcal{H}^B$, then a general
quantum state $\Psi\in\mathcal{H}$ looks like
\begin{equation}\label{E:2}
\Psi\ =\ \left[\begin{matrix} \Psi^F \\ \Psi^B \end{matrix}\right].
\end{equation}
Now let $M$ be a metric operator on $\mathcal{H}$ made up of the
identity operators $I^{FF}$ and $I^{BB}$ on
$\mathcal{H}^F$ and $\mathcal{H}^B$, respectively, as follows:
\begin{equation}\label{E:3}
M\ =\ \left[\begin{matrix} I^{FF} & 0 \\ 0 & -I^{BB}\end{matrix}
    \right]\ =\   M ^{-1}\ =\ M^\dagger.\,
\end{equation}
where the ``$\dagger$'' is the usual Hilbert space adjoint.
Then if $\Phi$ and $\Psi$ are two quantum states in $\mathcal{H}$
their inner product $(\Phi,\Psi)$ with respect to $M$
is taken to be
\begin{subequations}\label{E:4}
\begin{align}
(\Phi,\Psi)\ &=\ \Phi^\dagger M\Psi,\label{E:4a} \\
 &=\ \left[\begin{matrix} (\Phi^F)^\dagger & (\Phi^B)^\dagger
\end{matrix}\right]\left[\begin{matrix} I^{FF} &0\\ 0& -I^{BB}
\end{matrix}\right]\left[\begin{matrix} \Psi^F \label{E:4b}\\ \Psi^B
\end{matrix}\right] \\
&=\ (\Phi^F)^\dagger\Psi^F\ -\ (\Phi^B)^\dagger\Psi^B,\label{E:4c}\\
 &=\ \langle\Phi^F |\Psi^F\rangle_F\ -\ 
\langle\Phi^B | \Psi^B\rangle_B,\label{E:4d}
\end{align}
\end{subequations}
where $\langle\Phi^F |\Psi^F\rangle_F$  
and $\langle\Phi^B |\Psi^B\rangle_B$ are the 
Hilbert space inner products in Dirac notation.

     We presume a Schr\"odinger equation of the form
\begin{equation}\label{E:5}
i\hbar\frac{\partial\Psi}{\partial t}(t)\ =\ H(t)\Psi(t),
\end{equation}
where $t$ is the time, $\Psi(t)$ is as in \eqref{E:2}, and
$H(t)$ has the form
\begin{equation}\label{E:eq6}
H(t)\ =\  \left[\begin{matrix} H^{FF}(t) & H^{FB}(t)\\ H^{BF}(t)& H^{BB}(t)
\end{matrix}\right];
\end{equation}
note that $H^{FB}(t)$ is an interaction Hamiltonian, and 
performs a Hilbert space map
$\mathcal{H}^B$ into $\mathcal{H}^F$, while $H^{BF}$ performs
such a mapping in reverse.  As usual herein, the superscripts are to
be read from right to left:  the right superscript refers to
the operand, alias domain space, the left to the image space.
The Hamiltonian is to be self-adjoint with respect to the
metric operator $M$, that is
\begin{equation}\label{E:7}
H(t)\ =\ MH(t)^\dagger M,
\end{equation}
so that
\begin{subequations}\label{E:8}
\begin{align}
(H^{FF}(t))^\dagger \ &=\ H^{FF}(t),\label{E:8a}\\     
(H^{BB}(t))^\dagger \ &=\ H^{BB}(t),\label{E:8b}\\     
H^{BF}(t)\ &=\ -(H^{FB}(t))^\dagger.\label{E:8c}
\end{align}
\end{subequations}
In \eqref{E:8c}, the adjoint, denoted with a ``$^\dagger$'',
has the usual mathematical definition of the adjoint of a mapping of
one Hilbert space into another, that is
\begin{equation}\label{E:9}
\langle (H^{FB}(t))^\dagger \Psi^F(t)|\Psi^B(t)\rangle_B
\ =\ \langle \Psi^F(t)|H^{FB}(t)\Psi^B(t)\rangle_F
\end{equation}
defines a unique adjoint operator
such that the above is true for all admissible $\Psi^B(t)\in\mathcal{H}^B$
and $\Psi^F(t)\in\mathcal{H}^F$.

     Now suppose that the state vectors in \eqref{E:4a}
are both time dependent, and are both solutions of \eqref{E:5}.
We take the time derivative of their inner product:
\begin{subequations}
\begin{align}
i\hbar\frac{\partial}{\partial t} \bigl(\Phi(t),\Psi(t)\bigr)
\ &=\ i\hbar \frac{\partial \Phi^\dagger}{\partial t}(t) M\Psi(t)\,+\,
i\hbar \Phi(t)^\dagger M \frac{\partial\Psi}{\partial t}(t)\label{E:10a}\\
&=\ \Phi(t)^\dagger \bigl(-H^\dagger M\,+\, MH\bigr)\Psi(t)\label{E:10b}\\
&=\ 0,\label{E:10c}
\end{align}
\end{subequations}
where the last line results from \eqref{E:7} and \eqref{E:3}.
Therefore the inner product of any two solutions of the Schr\"odinger
equation is time independent;  in particular, the $M$-norm
of a time \-dependent quantum state, deconstructed as in
\eqref{E:2}, is conserved in time if $\Psi(t)$ satisfies 
\eqref{E:5}.  Let $[t_a,t_b]$, with $t_a<t_b$, be an
interval in time.  Then from \eqref{E:4d} we must have for $\Psi(t)$ that
\begin{equation}\label{E:11}
\langle \Psi^F(t_b)|\Psi^F(t_b)\rangle_F\,-\,
\langle \Psi^B(t_b)|\Psi^B(t_b)\rangle_B\ =\ 
\langle \Psi^F(t_a)|\Psi^F(t_a)\rangle_F\,-\,
\langle \Psi^B(t_a)|\Psi^B(t_a)\rangle_B.
\end{equation}
Rearranging, we have
\begin{equation}\label{E:12}
\langle \Psi^F(t_b)|\Psi^F(t_b)\rangle_F\,+\,
\langle \Psi^B(t_a)|\Psi^B(t_a)\rangle_B\ =\ 
\langle \Psi^F(t_a)|\Psi^F(t_a)\rangle_F\,+\,
\langle \Psi^B(t_b)|\Psi^B(t_b)\rangle_B.
\end{equation}
Both sides of \eqref{E:12} are positive definite.
This result suggests the following construction and interpretation,
as pictured in Fig.~1:
let the input and output to a physical process that takes place
for $t_a\leq t\leq t_b$
be the composite state vectors
\begin{subequations}\label{E:13}
\begin{align}
\Psi_{\rm in}(t_a,t_b)\ &=\ \left[\begin{matrix}
\Psi^F(t_a)\\ \Psi^B(t_b)\end{matrix}\right],\label{E:13a}\\
\Psi_{\rm out}(t_a,t_b)\ &=\ \left[\begin{matrix}
\Psi^F(t_b)\\ \Psi^B(t_a)\end{matrix}\right].\label{E:13b}
\end{align}
\end{subequations}
\begin{picture}(200,100)(-50,0)
\thicklines
\put(0,0){\dashbox(150,50){interaction zone}}
\put(150,0){\line(10,0){10}}
\put(150,50){\line(10,0){10}}
\put(165,-3){\shortstack{$t=t_a$}}
\put(165,47){\shortstack{$t=t_b$}}
\put(50,-20){\vector(0,10){20}}
\put(100,0){\vector(0,-10){20}}
\put(50,50){\vector(0,10){20}}
\put(100,70){\vector(0,-10){20}}
\put(40,-35){\shortstack{$\Psi^F(t_a)$}}
\put(90,-35){\shortstack{$\Psi^B(t_a)$}}
\put(40,75){\shortstack{$\Psi^F(t_b)$}}
\put(90,75){\shortstack{$\Psi^B(t_b)$}}
\put(220,10){\vector(0,1){20}}
\put(220,10){\vector(1,0){20}}
\put(215,35){\shortstack{time}}
\put(245,7){\shortstack{space}}
\end{picture}
\vskip 50pt
$\qquad\qquad$Figure 1.\ \ Wave function input/output scheme\\
\vskip 10pt\noindent
This makes physical sense:  the input to a physical process
in the time interval $[t_a,t_b]$
comprises the $F$-type signal at the beginning $t_a$
of the interval and the $B$-type signal at the end $t_b$, while the
output comprises the $F$-type signal at the end and
the $B$-type signal at the beginning of the time interval.

     An analogous imposition 
of boundary conditions applies in generalized classical dynamics:
for F-type particles we prescribe their positions and velocities at
the beginning of a time interval, while for B-type particles we should
specify their positions and velocities at the end of a time interval.
Correspondingly, it is an extension of observed properties of objects
in the real world that
an external perturbation on an otherwise isolated particle 
or system of particles 
affects only the future trajectories of F-type particles,
and would affect
only the past trajectories of B-type particles (neglecting
internal feedback interactions of a system in time).

    Note that \eqref{E:12} implies
that the positive-definite Hilbert-space norms
of the input and output states are equal:
\begin{equation}\label{E:14}
\Psi_{\rm out}(t_a,t_b)^\dagger\Psi_{\rm out}(t_a,t_b)
\ =\ \Psi_{\rm in}(t_a,t_b)^\dagger\Psi_{\rm in}(t_a,t_b).
\end{equation}
Therefore, nonnegative probabilities
are defined and conserved
overall, notwithstanding the indefinite metric $M$ introduced for
instantaneous state vectors.  The instantaneous $M$-norm
\eqref{E:4a} of a state vector represents the flow
of probability across the given $t=$constant surface:
$F$-type states' probability flows forward in time,
$B$-states' probability flows backward in time.   
The $S$-operator effects a transition from the input state to
the output state: 
\begin{equation}\label{E:15}
\Psi_{\rm out}(t_a,t_b)\ =\ S(t_a,t_b)\Psi_{\rm in}(t_a,t_b)
\end{equation}
and must therefore be unitary:
\begin{equation}\label{E:16}
S(t_a,t_b)^\dagger S(t_a,t_b)\ =\ I^{\mathcal{H}},
\end{equation}
where $I^{\mathcal{H}}$ is the identity operator in $\mathcal{H}$.
We now have a framework
that is globally, but not instantaneously, 
consistent with ordinary quantum mechanics, 
and makes a physically plausible way to describe temporally ``lumped''
processes in which both forward and backward evolution in time,
including transitions between the subspaces, can occur jointly.

     I have not been able to discover an instantaneous 
density matrix formalism and associated dynamics
that admits of a rearrangement to a mapping 
of input into output density matrices, 
thereby generalizing the state vector
transformation between \eqref{E:11} and \eqref{E:12}.
I infer that computation of the evolution of mixtures from input to
output must be subordinate to the treatment of the evolution of
each of an assembly of 
pure states, which can then be combined into mixtures at input and,
therefore, at output.  Entanglements between
F- and B-type states at input, or separately at output,
can be described in this formalism.

    We observe also that     
for instantaneous expectation values of a physical observable 
to be real, it must correspond to an operator self-adjoint
with respect to the metric $M$.  That is, let $Z(t)$ be a linear operator
in $\mathcal{H}$ for each time $t$ such that for all 
$\Psi(t)\in\mathcal{H}$ we have
\begin{equation}\label{E:17}
Z(t)^\dagger\ =\ MZ(t)M.
\end{equation}
If we define the expectation value of $Z(t)$ at time $t$ in the
state $\Psi(t)$ as
\begin{equation}\label{E;eq18}
\langle Z(t)\rangle_{\Psi(t)}\ =\ \Psi(t)^\dagger MZ(t)\Psi(t)
\ =\ \bigl(\Psi(t),Z(t)\Psi(t)\bigr),
\end{equation}
then $\langle Z(t)\rangle_{\Psi(t)}$ can easily be shown to be real;
in fact, 
\begin{equation}\label{E:19}
\bigl(\Psi(t)^\dagger MZ(t)\Psi(t)\bigr)^\ast\ =\ 
\Psi(t)^\dagger Z(t)^\dagger M^\dagger\Psi(t)\ =\ 
\Psi(t)^\dagger MZ(t)\Psi(t).
\end{equation}
We shall construe the expectation value of an operator at an instant
of time as the net flow of the corresponding quantity
(probability, energy, momentum, electric charge, baryon number, etc.)
across the complete spacelike surface.
This viewpoint therefore incorporates and generalizes the conventional
interpretation of expectation values, as now quantities can flow
backward as well as forward in time.  In particular, as noted 
by Pauli (\cite{R:Pauli1}, Sec. 2), operators with
only positive eigenvalues can have negative expectation values,
corresponding to the backward-in-time flow of positive ``stuff''.
We remark that, as discussed in \cite{R:Hahne1}, Sec.~3, for
time-independent, energy-conserving Hamiltonians
we shall assume the unperturbed $FF$ and $BB$ sub-Hamiltonians
both have positive eigenvalues, so that transitions between $F$
and $B$ states can be caused by time-independent $FB$ 
interactions.  (In this respect the present model differs
from the Klein-Gordon and Dirac wave equations, where
there are negative energy states in the unperturbed eigenvalue spectrum.
Those respective equations yield indefinite and positive-definite
conserved norms, respectively.)

     This concludes our review of the mathematical physics
underlying the proposed scheme.  
More details can be found in \cite{R:Hahne1}.
The remainder of this report is
organized as follows:  in Sec.~2, we shall discuss the physics 
of joint bidirectional motion in time, including $F\leftrightarrow B$
transitions, a possible structure for a theory of measurement
including the ``grandfather paradox'', and a suggested treatment
for gravitational radiation.  In Sec.~3 we shall
consider a simple model quantum field theory in which $F\leftrightarrow B$
transitions can occur, remark on the unavoidable cancellation
of vacuum energies in the model, and make a suggestion on the
possible generation of advanced gravitational waves.  
Sec.~4 addresses the question of the existence of magnetic
monopoles in $B$-type states (and not in $F$-type states).  Sec.~5
concludes the report with suggestions for further
research along these lines.


\section{2.  Some relevant physics}

     In this section we shall discuss some physics of
the direct sum construction, outline an (at present, undeveloped)
theory of measurement with reference to the ``grandfather paradox'',
and discuss how the present scheme might be made
consistent with gravitation in the form of Einstein's theory.
 
     The direct sum construction of \eqref{E:1} and \eqref{E:2}
has a different physical meaning from the direct product
construction (as, for example, in the wave function for two
or more particles) that is usual in combining subsystems
in quantum mechanics, or combining different quantum fields to
represent different families of physical particles.
We are saying that there is really only one physical system,
which has available quantum states within one of two antiparallel worlds.
Furthermore both of these worlds coexist on the same underlying
four-dimensional, pseudo-Riemannian, 
space-time manifold.  The latter assumption
imposes constraints on gravitational waves,
since in Einstein's geometric theory of gravitation, to which we shall try
to accommodate our argument,  the space-time manifold has
a metric structure that alone accounts for gravitational
phenomena.  Matter, and indeed vacuum energy, momentum, and pressure
of matter fields,
in both $F$ and $B$ states will be taken to be sources
of gravity.  We shall propose a decomposition of the energy-momentum-pressure
tensor into two tensor parts, one of which is coupled to
gravitation by radiating retarded (diverging as time advances) 
gravitational waves, the other by radiating advanced (converging)
gravitational waves.
We shall discuss these matters further later in this Section
and in Appendices A and B.

     The hypothesis of coexisting antiparallel worlds
of quantum states  will also have consequences when we try
to make a dynamical theory for a family of particles
that locally obeys Lorentz/Poincar\'e invariance
and can make transitions between the subspaces:
the two subspaces of quantum states will be constrained to resemble one
another to a certain degree beyond the assumption that they have in common the
gravity-generated four-dimensional space-time---see the discussion
in \cite{R:Hahne1}, Sec. 5.

     Let us now turn to the subject of measurements in the
extended quantum theory.  A possible setup is displayed in Fig.~2, below.
We shall not attempt a thorough quantitative analysis here.
This is the scenario:  Let $t_a<t_b<t_c<t_d$.
The specified overall input signal comprises
the direct sum of the vectors $\Psi^F(t_a)$ and $\Psi^B(t_d)$, 
the quantum output (to be determined) is the direct sum of $\Psi^F(t_d)$
and $\Psi^B(t_a)$.  In an intermediate time interval
$t_b\leq t\leq t_c$ measurements are performed on the system.
We presume, as shown in Fig.~2, that the measurements
are done separately on the $F$ and $B$ components of the state vector;
no reflections in the time dimension, that is, no
$F\leftrightarrow B$ transitions, are to be caused by the
measurement process.  Reflections can occur within interaction
zones 1 and 2, so that the ``trajectory'' of any particle
in the system has a 
possible feedback loop $1\rightarrow F\rightarrow 2\rightarrow B
\rightarrow 1$ or $2\rightarrow B\rightarrow 1\rightarrow F
\rightarrow 2$. 

     Suppose that the physical system consists of just one
particle:  The particle can traverse this loop
zero, one, two, three, \ldots times before exiting.  If the measurement
zones $F$ and $B$ do nothing more than record the particle's
passing through a respective zone, we will be able to count
the number of times the system has gone around the feedback
loop.  We will need to distinguish types of predictions,
according to whether or not we specify if, and how many,
feedback loops are undergone by the particle between
entry and exit of the overall time interval $[t_a,t_d]$.

\begin{picture}(200,250)(-50,0)
\thicklines
\put(0,0){\dashbox(150,50){interaction zone 1}}
\put(150,0){\line(1,0){10}}
\put(150,50){\line(1,0){10}}
\put(165,-3){\shortstack{$t=t_a$}}
\put(165,47){\shortstack{$t=t_b$}}
\put(50,-20){\vector(0,1){20}}
\put(100,0){\vector(0,-1){20}}
\put(50,50){\vector(0,1){30}}
\put(100,80){\vector(0,-1){30}}
\put(50,110){\vector(0,1){30}}
\put(100,140){\vector(0,-1){30}}
\put(40,-35){\shortstack{$\Psi^F(t_a)$}}
\put(90,-35){\shortstack{$\Psi^B(t_a)$}}
\put(40,215){\shortstack{$\Psi^F(t_d)$}}
\put(90,215){\shortstack{$\Psi^B(t_d)$}}
\put(220,10){\vector(0,1){20}}
\put(220,10){\vector(1,0){20}}
\put(215,35){\shortstack{time}}
\put(245,7){\shortstack{space}}
\put(0,80){\dashbox(70,30){meas.~zone F}}
\put(80,80){\dashbox(70,30){meas.~zone B}}
\put(0,140){\dashbox(150,50){interaction zone 2}}
\put(150,140){\line(1,0){10}}
\put(150,190){\line(1,0){10}}
\put(165,137){\shortstack{$t=t_c$}}
\put(165,187){\shortstack{$t=t_d$}}
\put(50,190){\vector(0,1){20}}
\put(100,210){\vector(0,-1){20}}
\put(15,60){\shortstack{$\Psi^F(t_b)$}}
\put(105,60){\shortstack{$\Psi^B(t_b)$}}
\put(15,120){\shortstack{$\Psi^F(t_c)$}}
\put(105,120){\shortstack{$\Psi^B(t_c)$}}
\end{picture}
\vskip 40pt
$\qquad\qquad$Figure 2.\ \ Intermediate measurements\\
\vskip 10pt

    Note also that this feedback property compels a revision
of the traditional picture of the ``collapse'' of a wave function
at the time of a measurement.  To be sure, 
it is plausible that if a measurement is performed
in the $F$ zone the wave function will collapse
in part between $\Psi^F(t_b)$ and $\Psi^F(t_c)$, 
although there is the possibility
that the particle crosses a given $t=$constant plane at
more than one spatial position in an $F$-type state due to feedback.
(An analogous partial collapse can occur in
the measurement that takes $\Psi^B(t_c)$ into $\Psi^B(t_b)$.)
Such a measurement also cannot
simultaneously wipe out the $B$ component of the wave
function, as a reflection can occur later in zone 2 that ensures
a nonzero chance of the system crossing measurement zone $B$
at the same time that the particle is detected in zone $F$.
A mathematical description of the input-to-output
of this system must be done in a self-consistent manner,
such that the information gleaned, and used or not used, in the intermediate
measurements is reckoned as part of the output.
I have made a tentative analysis of a classical analog to such a
feedback system \cite{R:Hahne2}, but have not yet undertaken a 
careful analysis of a quantum system of the type described.

     Assuming no intermediate measurements, 
a simple version of the grandfather
paradox (\cite{R:Nahin1}, \cite{R:Price1})  
might be encapsulated as follows with respect to
Fig.~2:  given a partial input
$\Psi^F(t_a)\neq 0$ and zero partial 
input at the later time ($\Psi^B(t_d)=0$),
a certain partial output $\Psi^F(t_d)$ will ensue.  Can one now
superimpose another partial input $\Psi^B(t_d)$ so that the
resultant partial output $\Psi^F(t_d)=0$?  It is plausible,
given suitable reflectivity properties of the interaction zones,
that the partial output generated at $t=t_d$
by the two partial inputs might be made to cancel exactly
by quantum interference.  This cancellation
could not plausibly occur in a classical probabilistic linear system
where the probabilities are all nonnegative.  (Some complex, nonlinear
classical probability 
scheme might do the job, however.)   It seems, therefore,
that realization of the grandfather paradox 
would be easier to effect (at least in a thought experiment)
when quantum, as opposed to classical, mechanics holds.

     We shall end this section with a preliminary discussion of the
role of gravitational waves.  It has been inferred from the
gradual winding down of the orbital motion of a binary star
(pulsar plus ordinary neutron star companion---see \cite{R:Weisberg1}) that
this system loses energy in the form of gravitational
radiation, in accord (to about $\pm 0.2\% $) with the 
quantitative predictions of Einstein's theory (\cite{R:Will1},
\cite{R:Wiki1}).
This outcome causes a problem in the present context:
as discussed in the next section, we will argue that
the expectation value of the vacuum energy due to the
$F$-type and $B$-type vacuum inevitably tend to cancel 
one another to high accuracy,
and that this cancellation has the observable consequence
that the cosmological constant is very small.  
This means that we are assuming that, while that
part of the world in $B$-type states is invisible
in the sense that it doesn't emit/absorb electromagnetic
waves of $F$ type, at least,  the part of the world in $B$
states does interact with the gravitational field, the same 
field that we, living in $F$-type states, interact with.
If gravitational waves are always of retarded type, however,
there would be an asymmetry of gravitational phenomena
with respect to time.  
Such a lopsidedness would tend 
to make the whole construct doubtful.  There is
a way around the problem, though, as we shall argue
\emph{ad hoc}, by assuming that the matter in certain quantum 
transitions 
emits (or rather from our viewpoint of view, absorbs)
gravitational radiation that
converges on the spacetime-localized transition.   Let me say that this
is the most promising physical test that I have been able to formulate
as yet to access the truth or falsity
of the hypothetical structure described herein.  This means that
there would be gravitational waves present in our universe
that have no visible source, as they are converging
on a future event that only emits light forward in time, at least
within the $F$-type world that we inhabit.

    Quantitatively, the hypothesis concerning gravitational
radiation looks as follows.  Suppose that 
$\kappa,\lambda,\mu,\nu=0,1,2,3$ are space-time coordinate indices.
Let $T^{\mu\nu}(\mathbf{r})$ be the operators representing the
energy-momentum-pressure tensor components, where the Hamiltonian $H$ is
([1], Eq.~(63))
\begin{equation}\label{E:20}
H\ =\ \iiint_{\rm{space}}d^3r\ T^{00}(\mathbf{r}).
\end{equation}
Einstein's field equations for gravity take the classical
energy-momentum-pressure tensor field to be the source of gravitation.
We shall construe this classical tensor field as expectation 
values of the quantum operators, that is,
\begin{equation}\label{E:21}
\langle T^{\mu\nu}\rangle(t,\mathbf{r})\ =\ \Psi(t)^\dagger
M T^{\mu\nu}(\mathbf{r})\Psi(t),
\end{equation}
where $\Psi(t)$ is the quantum state of the universe at time $t$
\emph{except gravity}---we are not attempting anything
more than a classical treatment of gravitational phenomena here.
Note also that the operators $T^{\mu\nu}(\mathbf{r})$ are presumed to 
incorporate the factor $16\pi G/c^4$.
The classical metric tensor field $g_{\mu\nu}(t,\mathbf{r})$
obeys a set of nonlinear partial differential equations;  the
source terms are those of \eqref{E:21}.  Given a suitable combination
of boundary conditions, 
and given the source terms,
we can solve the field equations for the metric tensor field.
(Gravity also affects the evolution of the sources,
so that a self-consistent procedure is required to get
physical solutions of matter plus gravity.)  

     At modest (say, galactic) scales space-time looks
pretty flat, so let us consider gravitational waves generated by some
source involving matter in $F$ or $B$ states on a background
flat spacetime, so that $g_{\mu\nu}=\rm{diag}(-c^2,1,1,1)$.
We represent these waves as a perturbation
$\delta g_{\mu\nu}(t,\mathbf{r})$ on the flat-space metric.  
To first order in perturbation
theory the $\delta g_{\mu\nu}$ must be linear functionals
of the source terms of \eqref{E:21}.  The functional operators that
take the sources into the generated gravitational waves are known as Green's
functions.  We have distinct choices for these Green's functions,
called retarded (left superscript $R$) and advanced (left superscript $A$):
\begin{equation}\label{E:22}
^R\mathcal{G}_{\mu\nu\kappa\lambda}(t,\mathbf{r};t',\mathbf{r}'),
\qquad
^A\mathcal{G}_{\mu\nu\kappa\lambda}(t,\mathbf{r};t',\mathbf{r}').
\end{equation}
The retarded Green's function $^R\mathcal{G}$ is zero for
$t<t'$ and says that a source generates outgoing gravitational waves,
while the advanced Green's function $^A\mathcal{G}$ is zero for
$t>t'$ and generates waves converging on the source as conventional
time advances.  Furthermore, let us 
assume that we can break up the source term
\eqref{E:21} into two parts, as follows: 
\begin{equation}\label{E:23}
\langle T^{\mu\nu}\rangle(t,\mathbf{r})\ =\ 
\langle ^R T^{\mu\nu}\rangle(t,\mathbf{r})\,+\,
\langle ^A T^{\mu\nu}\rangle(t,\mathbf{r}).
\end{equation}
We shall discuss in the Section 3 and in Appendix A how to effect
such a splitting of the matter field in a particular
quantum field theory model.
The upshot is that the perturbation of the metric tensor
is presumed to be
\begin{multline}\label{E:24}
\delta g_{\mu\nu}(t,\mathbf{r})\ =\ \sum_{\kappa,\lambda=0}^3
\iiiint_{\rm{spacetime}}\ dt'd^3r'
\biggl[{^R\mathcal{G}}_{\mu\nu\kappa\lambda}(t,\mathbf{r};t',\mathbf{r}')
\ \langle ^R T^{\kappa\lambda}\rangle(t',\mathbf{r}')
\\ \qquad\qquad
+\ ^A\mathcal{G}_{\mu\nu\kappa\lambda}(t,\mathbf{r};t',\mathbf{r}')
\ \langle ^A T^{\kappa\lambda}\rangle(t',\mathbf{r}')\biggr].
\end{multline}
Plausibly this formalism can be generalized to the case that the
``zeroth order'' background gravitational metric is some
cosmological or other nontrivial case, such that the gravitational
waves are generated as first-order perturbations on this background.  
I don't know how
to formulate this mixed retarded/advanced gravitational radiation
criterion as, say, boundary conditions
in the context of Einstein's full, nonlinear field equations.

     We shall discuss the possible detection of advanced
gravitational waves in Sec.~5.

\section{3.  Model quantum field theory}
     A model quantum field theory for Lorentz scalar, 
electrically neutral particles
was established in \cite{R:Hahne1}, Sec.~4.  The $FF$, $BB$, $FB$ and 
$BF$ interaction Hamiltonian densities were all presumed to be
proportional to $\phi(\mathbf{r})^4$---see \cite{R:Hahne1}, Eqs.~(62b)
and (63c).  We shall not review these basics here, but
elaborate on the results of \cite{R:Hahne1}, Secs.~ 4 and 5.

     The first question we want to address is whether any separation
of the expectation value of the energy-momentum-pressure tensor
into advanced and retarded sources---see \eqref{E:23} above---is
Lorentz invariant on the presumed flat background spacetime.
The Lorentz transformation properties of the full energy-momentum-pressure
tensor were asserted, based on unpublished work, in
\cite{R:Hahne1}, following Eq.~(95). 
The computation is nontrivial mainly for Lorentz ``boosts'', the generators
of which are given by \cite{R:Hahne1}, Eq.~(68d).
Second, we recapitulate the argument of \cite{R:Hahne1} that the
expectation value of the vacuum energy tends to cancel to
a high precision.  And third, we shall make a plausible speculation
concerning the nonrelativistic gravitational
interactions between, and dynamics of, two point particles.

     With respect to the Lorentz-invariant separation of
the energy-momentum-pressure tensor into two nontrivial parts:
There proves to be essentially one way to do this,
modulo fixed linear combinations of the constituent
tensors in the split.
Let us define projection operators in the space of quantum states:
\begin{subequations}\label{E:25}
\begin{align}
P^F\ &=\ \left[\begin{matrix} I^{FF} &0\\ 0 & 0\end{matrix}\right],
\label{E:25a}\\ 
P^B\ &=\ 
\left[\begin{matrix} 0 &0\\ 0 & I^{BB}\end{matrix}\right].\label{E:25b} 
\end{align}
\end{subequations}
Then if we define
\begin{subequations}\label{E:26}
\begin{align}
^D T^{\mu\nu}(\mathbf{r}) &= P^F T^{\mu\nu}(\mathbf{r}) P^F\,+\,
P^B T^{\mu\nu}(\mathbf{r}) P^B,\label{E:26a}\\
^N T^{\mu\nu}(\mathbf{r}) &= P^F T^{\mu\nu}(\mathbf{r}) P^B\,+\,
P^B T^{\mu\nu}(\mathbf{r}) P^F,\label{E:26b}
\end{align}
\end{subequations}
the separate expectation value arrays both prove to transform
as second rank 
symmetric tensors with respect to the Lie algebra generators
(\cite{R:Hahne1}, Eq.~(70)) of the Poincar\'e group---we shall omit
the details of the proof of this statement here.  Unlike the
summed tensor on the lhs of \eqref{E:23}, however, the
separate contributions on the rhs do not have zero four-divergence,
that is, do not represent conserved currents of energy and momentum
for the matter fields; therefore, we cannot use the these
entities directly in \eqref{E:24}.  What is needed is to 
subtract from the rhs of \eqref{E:26a} and add to the rhs of
\eqref{E:26b} a Lorentz-invariant quantity so that
both resulting tensors have zero four-divergence---the
latter entities can then be applied to \eqref{E:24}
using the advanced and retarded Green's functions  
derived from inverting d'Alembert's operator.
We exhibit some mathematical
steps along these lines in Appendix A.

   It was argued in \cite{R:Hahne1} that the net vacuum energy density
is forced by the dynamics to be
very small uniformly within a time interval
(see \cite{R:Hahne1}, Eq.~(94b)) 
when the discriminant of \cite{R:Hahne1}
Eq.~(75) is negative.  In physical terms, this means that the 
magnitude of the $F\leftrightarrow B$ energy  term  is greater than the
magnitude of the splitting between the energies of the
unperturbed $F$ and $B$ vacua.  When this inequality occurs,
the energy eigenvalues associated with the two vacuum states
are nonreal (and are complex conjugates of one another), so that
the quantum states are closed-channel states, such as 
often occur in nonrelativistic scattering theory.
The corresponding largeness of the off-diagonal $FB$ and $BF$
matrix elements of the Hamiltonian compared to
the splitting of $FF$ and $BB$ energies seems to mean that most
of the low-lying energy levels of matter in the universe are also closed
channels.  Only quantum states with relatively large energy
differences between the uncoupled $F$ and $B$ levels
therefore should correspond to open-channel, nontransient states
having real energy levels---it is intuitively plausible that our observed
universe comprises a mixture of nontransient, open-channel
energy states.  Therefore, there must be substantial differences
in the complexes of energy states occupied by the $F$-type
subworld and the $B$-type subworld, in order that the vacuum coupling
does not overwhelm the cosmology.  I have not yet undertaken
to do cosmological modeling by seeing which, if any, of the
input parameter values of \cite{R:Hahne1}, Eq.~(94b) are 
such that the present hypotheses can be chosen to be
consistent with what is known.

     Let us neglect all nongravitational interactions for the moment,
and consider what happens in the nonrelativistic, weak field
limit of the mutual interaction of two gravitating particles.
Consistent with Einstein's geometrization of gravity, 
we presume that the principle of equivalence 
of inertial and gravitational
mass holds in any circumstance.  
It is also plausible, on the basis of Einstein's theory, 
that Newton's equations hold to a first approximation
for the equations of motion of two particles of masses
$m_1$ and $m_2$, and positions $\mathbf{r}_1$ and $\mathbf{r}_2$,
whether they might
be both of F type, or both of B type, or of mixed type:
\begin{subequations}\label{E:27}
\begin{align}
m_1\frac{d^2\mathbf{r}_1}{dt^2}\ &=\ -Gm_1m_2\frac{\mathbf{r}_1
-\mathbf{r}_2}{|\mathbf{r}_1-\mathbf{r}_2|^3},\label{E:27a}\\
m_2\frac{d^2\mathbf{r}_2}{dt^2}\ &=\ -Gm_1m_2\frac{\mathbf{r}_2
-\mathbf{r}_1}{|\mathbf{r}_2-\mathbf{r}_1|^3}.\label{E:27b}
\end{align}
\end{subequations}
The above equations reduce to
\begin{subequations}\label{E:x27}
\begin{align}
\frac{d^2\mathbf{r}_1}{dt^2}\ &=\ -Gm_2\frac{\mathbf{r}_1
-\mathbf{r}_2}{|\mathbf{r}_1-\mathbf{r}_2|^3},\label{E:x27a}\\
\frac{d^2\mathbf{r}_2}{dt^2}\ &=\ -Gm_1\frac{\mathbf{r}_2
-\mathbf{r}_1}{|\mathbf{r}_2-\mathbf{r}_1|^3}.\label{E:x27b}
\end{align}
\end{subequations}
(See Appendix C for further development of the above dynamics.)
Particles of type F will have positive energy expectation values,
and hence positive inertial and gravitational masses;  
particles of type B will have
negative energy expectation values, and hence negative inertial 
and gravitational masses.
If (i) both $m_1>0$ and $m_2>0$, the particles accelerate toward one another;
if (ii) both $m_1<0$ and $m_2<0$, the particles accelerate away from 
one another.
If (iii) $m_1>0$ and $m_2<0$ the F-type particle 1 accelerates away from the
B-type particle 2,  while particle 2 accelerates toward 
particle 1.  If (iv) $m_1<0$ and $m_2>0$, the latter tendencies are
reversed.
Combining \eqref{E:x27a} and \eqref{E:x27b},
we infer that
\begin{equation}\label{E:y27}
\frac{d^2(\mathbf{r}_2-\mathbf{r}_1)}{dt^2}\ =\ -G(m_1+m_2)
\frac{\mathbf{r}_2-\mathbf{r}_1}{|\mathbf{r}_2-\mathbf{r}_1|^3}.
\end{equation}
If in case (iii) $m_1>-m_2>0$, the relative motion of the
particles exhibits acceleration toward one another, while if $-m_2>m_1>0$, the
relative motion shows acceleration away
from one another..  
I have not attempted to model the gravitohydrodynamics
of fluid mixtures of F-type and B-type matter.


\section{4. Magnetic monopoles}

     In this section we shall address the question 
of the existence of (at present, hypothetical) magnetic monopoles
\cite{R:Goldhaber1}, \cite{R:Cragie1}, \cite{R:Ficenec1}.
     
    Magnetic monopoles have not been convincingly detected
in experiments, albeit not for want of trying---see
\cite{R:Goldhaber1}, Sec.~II.
This asymmetry in electromagnetic physics
and in Maxwell's equations for the electromagnetic fields
has long been noted and questioned, despite the resulting mathematical 
convenience of being able to
define an electromagnetic vector potential field without
Dirac string singularities or other contrivances.  Dirac (\cite{R:Dirac1},
reproduced in \cite{R:Goldhaber1})
noted that the existence of just one magnetic monopole would ensure
the quantization of electric charge.  
Thomson (\cite{R:Thomson1}, reproduced
in \cite{R:Goldhaber1}) noted that there is angular momentum 
in the combined field of a fixed electric monopole and magnetic monopole:
if the electric charge is $e$, the magnetic charge
is $g$, and the vector from the electric charge to 
magnetic charge is $\mathbf{r}$,
then using the right-hand rule for converting pseudovectors to vectors 
we find, in Gaussian units, that
\begin{equation}\label{E:28}
\mathbf{L}\ =\ eg\mathbf{r}/(c|\mathbf{r}|).
\end{equation}
Note that the magnitude $|\mathbf{L}|$ of the angular momentum
is independent of the magnitude $|\mathbf{r}|$ of the separation distance
of the monopoles.  We believe that angular momentum in all forms is quantized,
so that in quantum theory, as observed by Saha \cite{R:Saha1},
\begin{equation}\label{E:29}
|\mathbf{L}|\ =\ \lambda\hbar,
\end{equation}
where it is uncertain whether to choose $\lambda=1/2$, 
$\lambda=1$, or possibly some
other such value.  Dirac \cite{R:Dirac1} recommended $\lambda=1/2$, 
while Schwinger
\cite{R:Schwinger1} argued for an even integral value, say $\lambda=2$;
more recent works (see papers reproduced in \cite{R:Goldhaber1}) come down
in favor of $\lambda=1/2$ or $\lambda=1$.

     Provided that electric charges emit retarded electromagnetic signals
and magnetic charges emit advanced signals, \eqref{E:28} has an
(at least approximate)
relativistic generalization.  Let a classical electric charge have the
spacetime trajectory given by $(t_e,\mathbf{r}_e(t_e))$;  then the forward
light cone from each point on this trajectory intercepts the spacetime
trajectory of the magnetic monopole at a unique point $(t_m(t_e),
\mathbf{r}_m(t_m(t_e)))$, such that the four-vector from electric pole
to magnetic pole is a null vector.  The two poles communicate, as it were,
back and forth along this one-parameter family of null vectors.  
Let us now introduce some group
representation theory.  The so-called ``little group'', 
in the sense of Wigner (\cite{R:Wigner1}, see also \cite{R:Mackey1})  
as the subgroup of the
Poincar\'e group that leaves this null vector invariant, is known
to be isomorphic to the three-parameter group of translations and
rotations in the Euclidean plane.  In fact, let us 
first translate, rotate, and boost
a corresponding pair of trajectory points so that 
the electric pole is at $(0,0,0,0)$ and the magnetic pole at the remove
(a/c,0,0,a), the latter being a null vector.  
Then clearly rotations in the $xy$ plane leave the two charges
unmoved.  There is also another, two-parameter, family of joint
rotations and boosts that leaves the two positions unchanged,
but we shall not display those transformations here.  Rotations in the
$xy$ plane correspond to rotations around the $z$ axis, 
along which the null vector
has its spatial projection.  If we quantize the angular momentum conjugate to
rotations around the $z$-axis, we achieve a relativistic generalization
of the quantization entailed in \eqref{E:28} and \eqref{E:29}.

     As noted, Thomson showed that the angular momentum associated
with the circulating electromagnetic field momentum for
a spatially fixed electric-magnetic monopole pair is given by
\eqref{E:28}.  However, we are assuming herein that
there are no direct electromagnetic interactions between particles
in $F$ states and those in $B$ states;  hence the magnetic field generated
by a magnetic monopole in a $B$ state would live entirely in the
$B$ subworld, and similarly for the electric
field of an electric monopole in an $F$ state.
\emph{I propose now to drop the assumption that the
relative interaction between electric and 
magnetic charges and currents results from direct electromagnetic
interaction}, and to
take as \emph{ad hoc}, tentative physical axioms
that magnetic monopoles exist in, and only in, $B$-type
states and that each electric-magnetic monopole pair
has a relative, and relativistic, angular momentum
of the type described in the previous paragraph.  
Although 
the physical electromagnetic fields associated with the two charges,
in particular, the radiation fields associated with an accelerating
charge, 
do not interact directly with the other charge in such a pair,
the two charges must nevertheless interact:  their
movements must be correlated so that this axial angular momentum
plus any relativistic orbital angular momentum associated
with the particles' motion is conserved componentwise---there
are six components of angular momentum in a four-dimensional
spacetime (the three rotations plus three Lorentz boosts
make up the 
six-parameter ``rotation'' group in Minkowski spacetime).
The product of elementary units of electric and magnetic charge
would therefore be quantized,
and close encounters of electric and magnetic
monopoles would result in mutual scatterings.  
I do not presently know how to
formulate this hypothetical interaction in terms of quantum
field theory, such that the presumed 
angular momentum conservation and charge quantization
are natural consequences entailed by the mathematics.
[Note added:  PHYSICS TODAY of July, 2006 \cite{R:PT1} 
contains a news article on a recent search, with a negative outcome, for
magnetic monopoles at the Fermilab Tevatron.] 


\section{5.  Further research}
     The picture proposed herein of two antiparallel but somehow
disjoint spacetimes may be convenient for visualization, but is
dispensible.  As remarked above, 
a picture more in accord with quantum mechanics
and with Einstein's geometric theory of gravitation is that there is just
one four-dimensional spacetime, within which there inhere
two families of quantum states with their dynamical time
evolutions being oppositely directed, and no direct
electromagnetic interactions between the two subfamilies.
The picture of just two families of quantum states,
apart from dynamical considerations, can be construed as an
extreme specialization of the much-discussed scenario of extra
continuous dimensions to spacetime.  In the latter case there would
be infinitely many families of quantum states (countably infinite
if spacetime is compact in the extra dimensions).  I believe
this proliferation of Lorentz-congruent families of particles 
to be excessive and unphysical;  in the present scenario any ``extraness''
to spacetime or to particle families is discrete, finite, and small.

     It is a consequence of the arguments leading to \eqref{E:24}
that certain transitions between F and B states would give rise to
advanced gravitational waves (and in our subworld, retarded electromagnetic
waves).  The waves, to be observable by us, would necessarily arise
from intense astrophysical events.
This gravitational radiation would
appear to us to be converging on the future explosion/implosion,
and if it exists, should by our hypotheses be 
detectable with gravitational wave detectors.  Two such detectors are
under development or in the early stages of testing---they
are known by the acronyms LISA \cite{R:Stroeer1} and LIGO
\cite{R:Abbott1}, respectively Laser Interferometer Space
Antenna and Laser Interferometer Gravitational Wave Observatory.
Both of these detectors have directional capability.  
The spherical waves associated with both remote 
astrophysical events would locally
resemble plane waves, so it would not be possible to distinguish
exploding from imploding waves directly.  
What might be discovered, however, is gravitational waves
that have no visible source in the heavens, that is such that there are
no light signals plausibly associated with a gravitational wave
by having the same
direction and approximately the same arrival time.  Such waves
could, despite our normal guess that a concealed explosion
is the source, actually be converging on an implosion in the
diametrically opposite direction in space and time.
(Admittedly, this search for invisible-future-sourced gravitational
waves could be confused by the presence of invisible-past-sourced
gravitational waves, which can arise from violent
events within the $B$-subworld, 
for according to \eqref{E:26a} and \eqref{E:24},
such events give rise to retarded gravitational waves;
perhaps the detailed structure of the signals plus mathematical models
could make it possible to distinguish these two sorts of waves.)
I propose, therefore, that gravitational wave detectors spend
a modest effort in searching for gravitational waves
propagating with directions chosen more or less at random,
in particular in directions that might be unpromising
in a search for explosions in the $F$ world.

     In the absence of more detailed modeling, I cannot guess
how concentrations of F-type matter (stars, galaxies, cosmos) might interact
with concentrations of B-type matter, so as to provide
scenarios in which the latter might show their existence to us.
The hypothesized interaction  
between magnetic and electric monopoles
might lead to observable effects.  A moving magnetic monopole
in a $B$ state passing through ordinary matter would to an extent simulate
the effects of a moving monopole in an $F$ state, not through its
magnetic field but through the presumed axial angular momentum
between the two kinds of monopole. Such a search for moving $F$-type 
magnetic monopoles
could just as well be applied to search for $B$-type
monopoles;  a scattering would occur, but no magnetic monopole field
would be present, and the monopole would enter and exit the
experiment leaving no other trace of its presence.  Although I have not
analyzed this scenario in detail, 
it is plausible that a $B$-type magnetic monopole
would gain energy (in our positive time direction) when passing 
through $F$-type matter, as its second law of thermodynamics is operating in 
reverse.  In this connection, note that the future behavior of a
$B$-type particle is to a great extent predetermined by the presumed
boundary conditions on the problem:  perturbations
or measurements on the particle, will change its past, but not (apart from
complex feedback-loop mechanisms) its future, evolution.
Therefore ``trapping'' a magnetic monopole for long-term
observation would be impractical, as random forces on it would
likely be associated with anti-damping of its motion and
its eventual escape as time goes forward.

     The model described above,
of exactly one $F$-type world and exactly one $B$-type world
coexisting on the same 
four-dimensional space-time manifold, admits of straightforward
generalizations to larger (or lesser) numbers of parallel/antiparallel worlds.
A more complicated case is specified and discussed briefly
in Appendix B.
We have in the above kept to the minimum nontrivial numbers
in both cases in order to exhibit the beginnings of a plausible
physical theory to describe phenomena involving both forward
and reverse causation in time.  In this connection---see Appendix B---a 
second parallel $F$-type (as opposed
to an antiparallel $B$-type) subworld might contain matter that
interacts 
mainly through (straightforwardly
attractive) gravitation with matter in our subworld.  
The proposals for dark matter in 
galaxies \cite{R:Ashman1} and in galaxy clusters
in the form of weakly interacting
massive particles (WIMPs)---see e.~g.~\cite{R:Feng1}---resemble
this parallel-world scenario, except that the latter entails
a subworld unto itself, with possibly complex structure and mutual interactions
among its constituents.  I do not presently know how
to reconcile either of these hypotheses with the apparent absence
of WIMPs or invisible $F$-type particles in stars.

     Much theoretical
work remains to be done along these lines, as is made explicit in the
text above, to develop
the extent and show the consistency of the model.
We have described some
possible ways to subject these hypotheses to experimental
confirmation or disconfirmation. Perhaps more, and more feasible,
tests will emerge from further study.

\section{Appendix A:  Green's functions}

        The linearized theory of gravity is presented in \cite{R:Misner1},
Ch.~18---see particularly Eqs.~(18.7)---(18.8c) and Box 18.2.   
A consistent formalism of the desired type \eqref{E:24}, above, 
is obtained if we (i) break up the energy-momentum-pressure tensor
into two Lorentz-invariant summands that each have zero four-divergence, 
(ii) break up the perturbation on the background Minkowski metric
into two summands, each of which satisfies the Lorentz condition
(\cite{R:Misner1}, Eq.~(18.8a)), (iii) use Eq.~(18.8b)
for each corresponding summand in the source and the metric
perturbation, and (iv) apply \eqref{E:24}.  In the following, 
$x=(x^0,\mathbf{x})$, $x^0=ct$, Greek indices range from 0 to 3,
and the summation convention applies.  We use the Minkowski metric 
$\eta_{\mu\nu}=\rm{diag}(-1,1,1,1)$ to be consistent with \cite{R:Misner1}.

     The Green's function for d'Alembert's operator satisfies
\begin{equation}\label{A1}
-\eta^{\mu\nu}\frac{\partial^2}{\partial x^\mu x^\nu}
G^{\pm}(x;y)\ =\ \delta^4(x-y).
\end{equation}
Then we have
\begin{equation}\label{A2}
G^{\pm}(x;y)\ =\ 
\frac{-1}{(2\pi)^{4}}\iiiint d^4k\exp[-ik^0(x^0-y^0)+i\mathbf{k}
\cdot(\mathbf{x}-\mathbf{y})][(k^0\pm i\epsilon)^2-\mathbf{k}\cdot\mathbf{k}]
^{-1}
\end{equation}
where $\epsilon$ is the usual small positive quantity.  The
upper sign gives rise to 
a retarded Green's function, the lower sign an advanced
Green's function.  We then take for the Green's functions
of \eqref{E:22} the following:
\begin{subequations}\label{A3}
\begin{align}
^R\mathcal{G}(x;y)_{\mu\nu\kappa\lambda}\ &=\ 
\eta_{\mu\kappa}\, \eta_{\nu\lambda} G^+(x;y),\\
^A\mathcal{G}(x;y)_{\mu\nu\kappa\lambda}\ &=\ 
\eta_{\mu\kappa}\, \eta_{\nu\lambda} G^-(x;y).
\end{align}
\end{subequations}
Let us now define
\begin{multline}\label{A4}
\langle\Delta T^{\mu\nu}\rangle(x)\ =\ \eta^{\mu\kappa}
\frac{\partial}{\partial x^\kappa}\iiiint d^4y\, G^+(x,y)
\frac{\partial}{\partial y^\lambda}\langle ^NT^{\lambda\nu}
\rangle (y)\\
+ \eta^{\nu\kappa}
\frac{\partial}{\partial x^\kappa}\iiiint d^4y\, G^+(x,y)
\frac{\partial}{\partial y^\lambda}\langle ^NT^{\mu\lambda}\rangle(y)\\
+\eta^{\mu\kappa}\eta^{\nu\lambda}\frac{\partial^2}
{\partial x^\kappa \partial x^\lambda}\iiiint d^4y
\iiiint d^4z\, G^+(x,y)G^+(y,z)
\frac{\partial^2}{\partial z^\rho\partial z^\sigma}
\langle ^NT^{\rho\sigma}\rangle (z).
\end{multline}
Then if we take
\begin{subequations}\label{A5}
\begin{align}
\langle ^RT^{\mu\nu}\rangle(x)\ &=\ \langle ^DT^{\mu\nu}\rangle(x)
-\langle\Delta T^{\mu\nu}\rangle(x),\label{A5a}\\
\langle ^AT^{\mu\nu}\rangle(x)\ &=\ \langle ^NT^{\mu\nu}\rangle(x)
+\langle\Delta T^{\mu\nu}\rangle(x),\label{A5b}
\end{align}
\end{subequations}
both rhs's have zero net four-divergence and can be applied
to \eqref{E:24} with \eqref{A3}.  Note that the choice of,
among other possibilities, $G^+$
on the rhs of \eqref{A4} has the physical property of 
concentrating sources of advanced gravitational waves
in the future of an astrophysical event that generates
a nonzero $\langle ^NT^{\mu\nu}\rangle$.
The mathematics of \eqref{E:24} with \eqref{A4} entails the
convolution of two or three Green's functions;  I do not know
whether or not this procedure is mathematically justifiable.

     A possible way, which I have not investigated, 
to avoid the mathematically questionable
procedure entailed by \eqref{A4} might be to try to construct
less singular, but not Lorentz invariant, forms of
$\langle \Delta T^{\mu\nu}\rangle(x)$,
and redefine the generators of Lorentz boosts (\cite{R:Hahne1},
Eq.~(68c)) to include infinitesimal gauge transformations 
(\cite{R:Misner1}, Box 18.2B) so that
the commutators and transformation properties are restored.

\section{Appendix B:  Further discussion}

      We consider now that spacetime supports the direct sum
of three families of quantized fields:  (1) the 
forward-evolving fields (denoted $F_1$)
corresponding to conventional nondark matter as it is 
displayed to us by nongravitational as well as by
attractive gravitational interactions, (2) a second
family of forward-evolving fields (denoted $F_2$), the presence of which
is detectable primarily through its attractive gravitational
interactions with matter in our subworld, and (3) a third family
of, in this case, 
backward-evolving fields (denoted $B$) that also makes its presence
in space-time known primarily through attractive/repulsive gravitational
interactions with matter in the $F_1$ and $F_2$ subworlds---see
\eqref{E:27}, et seq.  We advance the conjecture that
the second and third families are candidates for dark matter and dark
energy, respectively.

      A natural question is, how can one distinguish physically between
taking the direct sum of, as contrasted with the direct product of,
the fields $F_1$ and $F_2$ (and possibly $B$)
as the space of states of the system?   I think
an irreconcilable difference is that in the direct sum, there is
more than one vacuum state, exactly one for each subworld.
A general vacuum state is thus a superposition or mixture of this
presumably small number of separate
vacua, where the time evolution of the
superposition or mixture is controlled by the dynamics
and the boundary conditions, as in \cite{R:Hahne1}, Eq.\ (80), et seq.

\begin{picture}(300,500)(0,0)
\thicklines
\put(0,0){\line(1,0){200}}
\put(50,100){\line(1,0){200}}
\put(0,0){\line(1,2){50}}
\put(200,0){\line(1,2){50}}
\put(75,0){\line(1,1){100}}
\put(175,0){\line(-1,1){100}}
\put(100,15){\shortstack{cone}}
\put(102,75){\shortstack{light}}
\put(225,0){\shortstack{Backward (B) sub-spacetime}}
\put(275,90){\vector(1,0){20}}
\put(275,90){\vector(0,-1){20}}
\put(305,85){\shortstack{space}}
\put(250,50){\shortstack{dynamical time evol.}}
\put(125,50){\circle*{8}}
\put(135,46){\shortstack{(t,x,y,z)}}
\put(0,150){\line(1,0){200}}
\put(50,250){\line(1,0){200}}
\put(0,150){\line(1,2){50}}
\put(200,150){\line(1,2){50}}
\put(75,150){\line(1,1){100}}
\put(175,150){\line(-1,1){100}}
\put(100,165){\shortstack{cone}}
\put(102,225){\shortstack{light}}
\put(225,150){\shortstack{Forward ($F_2$) sub-spacetime}}
\put(275,195){\vector(1,0){20}}
\put(275,195){\vector(0,1){20}}
\put(305,190){\shortstack{space}}
\put(250,225){\shortstack{dynamical time evol.}}
\put(125,200){\circle*{8}}
\put(135,196){\shortstack{(t,x,y,z)}}
\put(0,300){\line(1,0){200}}
\put(50,400){\line(1,0){200}}
\put(0,300){\line(1,2){50}}
\put(200,300){\line(1,2){50}}
\put(75,300){\line(1,1){100}}
\put(175,300){\line(-1,1){100}}
\put(100,315){\shortstack{cone}}
\put(102,375){\shortstack{light}}
\put(225,300){\shortstack{Forward ($F_1$) sub-spacetime}}
\put(275,345){\vector(1,0){20}}
\put(275,345){\vector(0,1){20}}
\put(305,340){\shortstack{space}}
\put(250,375){\shortstack{dynamical time evol.}}
\put(125,350){\circle*{8}}
\put(135,346){\shortstack{(t,x,y,z)}}
\end{picture}
\vskip 10pt
$\qquad\qquad$Figure 3.\ \ Parallel/antiparallel subworlds
\vskip 10pt     
     The above scenario therefore differs from the hypothesis of the existence
of WIMPs in that there is presumed to be a second vacuum state associated
with the $F_2$ subworld, and also associates the existence of
dark energy with a third vacuum state, that belonging to the B subworld.

\section{Appendix C:  Newtonian gravitational 
dynamics with a positive and a negative mass particle}

(Note:  The material in this Section augments the paper published
in \cite{R:Hahne3}.)  

     Suppose that, in \eqref{E:x27}, $m_1>0$ and $m_2<0$, so that,
respectively, these particles belong to the $F$ and $B$
subspaces. Then in establishing boundary value problems for
the Newtonian equations of motion in a time interval, 
we take it as a physical axiom that we must specify, respectively, 
the initial
and the final positions and velocities of the $F$- and $B$-type particle.
Suppose that $t_a\leq t\leq t_b$, and let it be given that
\begin{alignat}{2}\label{E:C1}
{\bf r}_1(t_a)\ &=\ {\bf r}_{1a}, &
\qquad\dot{\bf r}_1(t_a)\ &=\ {\bf u}_{1a},\\
{\bf r}_2(t_b)\ &=\ {\bf r}_{2b}, &
\qquad\dot{\bf{r}}_2(t_b)\ &=\ {\bf u}_{2b}.
\end{alignat}
The desired functions then are the solutions of the coupled
integral equations
\begin{subequations}\label{E:C2}
\begin{align}
{\bf r}_1(t)&={\bf r}_{1a}+{\bf u}_{1a}(t-t_a)+Gm_2\int_{t_a}^{t} dt'(t-t')
\frac{[{\bf r}_2(t')-{\bf r}_1(t')]}{|{\bf r}_2(t')-{\bf r}_1(t')|^3},\\
{\bf r}_2(t)&={\bf r}_{2b}-{\bf u}_{2b}(t_b-t)-Gm_1\int_{t}^{t_b} dt'(t'-t)
\frac{[{\bf r}_2(t')-{\bf r}_1(t')]}{|{\bf r}_2(t')-{\bf r}_1(t')|^3}.
\end{align}
\end{subequations}
If we define
\begin{equation}\label{E:C3}
\Vec{\Delta}(t_a,t_b)=\int_{t_a}^{t_b}dt'
\frac{[{\bf r}_2(t')-{\bf r}_1(t')]}{|{\bf r}_2(t')-{\bf r}_1(t')|^3},
\end{equation}
then
\begin{subequations}\label{E:C4}
\begin{align}
\dot{\bf{r}}_1(t_b)\ &=\ {\bf u}_{1a}+Gm_2\Vec{\Delta}(t_a,t_b),\\
\dot{\bf{r}}_2(t_a)\ &=\ {\bf u}_{2b}+Gm_1\Vec{\Delta}(t_a,t_b).
\end{align}
\end{subequations}

Since energy is conserved, if we take $t=t_a$ well before,
and $t=t_b$ well after, the time of closest approach,
we have equality of the kinetic energies with small potential energy:
\begin{equation}\label{E:C5}
(m_1/2)|\dot{\bf{r}}_1(t_b)|^2 + (m_2/2)|\dot{\bf{r}}_2(t_b)|^2
\ \approx\ (m_1/2)|\dot{\bf{r}}_1(t_a)|^2 + (m_2/2)|\dot{\bf{r}}_2(t_a)|^2.
\end{equation}
Since $m_2<0$, the above implies
\begin{equation}\label{E:C6}
(m_1/2)|\dot{\bf{r}}_1(t_b)|^2 + (|m_2|/2)|\dot{\bf{r}}_2(t_a)|^2
\ \approx\ (m_1/2)|\dot{\bf{r}}_1(t_a)|^2 + (|m_2|/2)|\dot{\bf{r}}_2(t_b)|^2
\end{equation}
that is, the sum of the absolute values of the output kinetic energies is
equal to the sum of the absolute values of the input kinetic 
energies---runaway solutions are impossible so long as the
interparticle distance is bounded away from zero.

\vskip 10pt
\section{Acknowledgement}
This report was published with the proceedings
of the Symposium on ``Frontiers of Time: Reverse Causation---Experiment
and Theory'', held as part of the AAAS Pacific Division Meeting,
University of San Diego, June 18--22, 2006.   I thank Prof. Daniel
Sheehan of the Department of Physics, University of San Diego, 
for convening the Symposium and for shepherding the publication
of the proceedings.

\vskip 10pt
\begin{thebibliography} {10}
\bibitem{R:Hahne1}
G.~E.~Hahne.
\newblock{\emph{J.~Phys.~}{\bf A35}, 7101 (2002).
Online in a slightly augmented version at arXiv:quant-ph/0404103.}
\bibitem{R:Pauli1}
W.~Pauli.
\newblock{Rev.~Mod.~Phys.~{\bf 15}, 175 (1943).}
\bibitem{R:Hahne2}
G.~E.~Hahne.
\newblock{\emph{Unpublished work}.}
\bibitem{R:Nahin1}
P.~J.~Nahin.
\newblock{\emph{Time Machines} (AIP Press, 1999), 2nd edition, passim.}
\bibitem{R:Price1}
H.~Price.
\newblock{\emph{Time's Arrow and Archimedes' Point}
(Oxford U. Press, 1996), p.~170, et seq.}
\bibitem{R:Weisberg1}
Joel M.~Weisberg and Joseph H.~Taylor.
\newblock{In \emph{Binary Radio Pulsars}, 
edited by F.~A.~Rasio and I.~H.~Stairs
(ASP Conference Series,
Vol.~328, 2004).  Online at arXiv:astro-ph/0407149 v1.}
\bibitem{R:Will1}
Clifford M.~Will.
\newblock{\emph{The Confrontation between General Relativity and
Experiment}, Living Rev.~Relativity {\bf 9}, (2006), 3
(cited June 1, 2006),
URL:  http://www.livingreviews.org/lrr-2006-3 .}
\bibitem{R:Wiki1}
Author/editor not given.
\newblock{\emph{Gravitational Radiation},
Wikipedia (cited June 1, 2006),
URL:  http://en.wikipedia.org/wiki/Gravitational\_radiation .}
\bibitem{R:Watson1}
G.~S.~Watson. 
\newblock{\emph{An Exposition on Inflationary Cosmology.}
Online at arXiv:astro-ph/0005003 v2.}
\bibitem{R:Carroll1}
Sean M.~Carroll.
\newblock{\emph{The Cosmological Constant}, Living Rev.~Relativity
{\bf 4},1 (cited June 1, 2006), URL:  
http://www.livingreviews.org/lrr-2001-1;  
also online at arXiv:astro-ph/0004075 v2.}
\bibitem{R:Goldhaber1}
Alfred S.~Goldhaber and W.~Peter Trower, eds.
\newblock{\emph{Magnetic Monopoles} (Amer.~Assoc.~of Phys.~Teachers,
College Park, MD, 1990).}
\bibitem{R:Cragie1}
N.~S.~Cragie, P.~Goddard, and W.~Nahm, eds.
\newblock{\emph{Monopoles in Quantum Field Theory},
(World Scientific, Singapore, 1982).}
\bibitem{R:Ficenec1}
John R.~Ficenec and Vigdor L.~Teplitz.
\newblock{\emph{Magnetic Charges}, In \emph{Electromagnetism,
paths to Research}, edited by Doris Teplitz (Plenum Press, New York, 1982).}
\bibitem{R:Dirac1}
P. A. M. Dirac.  
\newblock{Proc.~R.~Soc.~London Ser.~{\bf A133}, 60 (1931).}
\bibitem{R:Thomson1}
J.~J.~Thomson.
\newblock{Phil.~Mag.~{\bf 8}, 331 (1904).}
\bibitem{R:Saha1}
M.~N.~Saha.
\newblock{Ind.~J.~Phys.~{\bf 10}, 145 (1936); Phys.~Rev.~{\bf 75},
1968 (1949), reproduced in \cite{R:Goldhaber1}.}
\bibitem{R:Schwinger1}
J.~Schwinger.
\newblock{Phys.~Rev.~{\bf 144}, 1087 (1965).}
\bibitem{R:Wigner1}
E.~Wigner.
\newblock{Ann.~Math.~{\bf 40}, 149 (1939).}
\bibitem{R:Mackey1}
G.~W.~Mackey.
\newblock{\emph{Induced Representations of Groups and Quantum Mechanics}
(Benjamin, New York, 1968).}
\bibitem{R:PT1}
B.~Schwarzschild.
\newblock{\emph{Search for magnetic monopoles at the 
Tevatron sets new upper limit on their production},
Physics Today  {\bf 59/7}, 16--18 (July, 2006).}
\bibitem{R:Stroeer1}
A.~Stroeer and A.~Vecchio.
\newblock{\emph{The LISA verification binaries}. 
Online at arXiv:astro-ph/0605227 v1.}
\bibitem{R:Abbott1}
B.~Abbott, et al. 
\newblock{\emph{Coherent searches for periodic gravitational 
waves from unknown isolated sources and Scorpius X-1:
results from the second LIGO run}.  Online at
arXiv:gr-qc/0605028 v1.}
\bibitem{R:Ashman1}
K.~M.~Ashman.
\newblock{Publ.~Astron.~Soc.~Pacific {\bf 104},1109 (1992).}
\bibitem{R:Feng1}
J.~L.~Feng, A.~Rajaraman, and F.~Takayama.
\newblock{Phys.~Rev.~Lett.~{\bf 91}, 011302 (2003).}
\bibitem{R:Misner1}
C.~W.~Misner, K.~S.~Thorne, and J.~A.~Wheeler.
\newblock{\emph{Gravitation}
(W.~H.~Freeman, San Francisco, 1973).}
\bibitem{R:Hahne3}
G.~E.~Hahne, in
\newblock{\emph{Frontiers of Time}, edited by D.~P.~Sheehan
(AIP, Melville, New York, 2006), pp.~44--61.}
\end{thebibliography}
\end{document}